# Temperature dependent photoluminescence from single-walled carbon nanotubes

J. Lefebvre[*], P. Finnie[*], Y. Homma[§]

[*]*Institute for Microstructural Sciences, National Research Council, Montreal Road, Ottawa, Ontario, K1A OR6, Canada.*

[§]*NTT Basic Research Laboratories, Nippon Telegraph and Telephone Corporation, 3-1 Morinosato-Wakamiya, Atsugi, Kanagawa 243-0198, Japan*

**Photoluminescence (PL) and photoluminescence excitation (PLE) spectroscopy of pillar-suspended single-walled carbon nanotubes has been measured for temperatures between 300 K and 5 K. The atmospheric environment strongly affects the low temperature luminescence. The PL intensity is quenched at temperatures below ~40 K for nanotubes in high vacuum, while nanotubes in helium ambient remain luminescent. The PL peak emission energy is only very weakly dependent on temperature, with a species dependent blueshift upon cooling corresponding to a relative shift in bandgap of $-3 \times 10^{-5}$ K$^{-1}$ or less. The integrated peak intensities change by only a factor of two, with linewidths showing a moderate temperature dependence. In PLE, the second absorption peak energy ($E_{22}$) is also only weakly temperature dependent, with no significant shift and only a limited reduction in linewidth upon cooling to 20K. In addition to the previously assigned nanotube PL peaks seen at room temperature, at least two distinct new classes of PL peaks are observed at cryogenic temperatures.**

## I. INTRODUCTION

Recently, we reported photoluminescence (PL) from single-walled carbon nanotubes (SWNTs) suspended between pillars above silicon substrates.[1] Suspension reduces environmental interactions and prevents quenching of luminescence, which can occur when nanotubes become bundled or contact substrates and other materials. Earlier measurements made on a solution of micelle-encapsulated SWNTs indicated that the PL arises from electron-hole recombination at the lowest subbands of semiconducting nanotubes (at energy $E_{11}$, where the indices stand for the first conduction and valence subbands in a single particle picture of the band structure).[2] The band structure of SWNTs has been revealed by photoluminescence excitation (PLE) spectroscopy, where the PL intensity is strongest whenever the excitation energy is resonant with higher order subbands (at energy $E_{22}$, $E_{33}$, …).[3,4,5,6]

In addition to ensemble measurements, PL from individual nanotubes can be readily measured on these pillar samples.[7] Individual nanotube PL has also been detected from micelle-encapsulated nanotube solutions dispersed on microscope slides.[8] Compared to ensemble measurements, spectra from individual nanotubes are greatly simplified, and in many ways clearly reveal the one-dimensional nature of SWNTs: polarized emission and absorption, and in particular for pillar-suspended nanotubes, asymmetric lineshapes with sub-$k_BT$ linewidths at room temperature ($k_BT$ is the thermal energy, where $k_B$ is the Boltzmann's constant).

Pillar-suspended nanotubes are well suited to temperature dependent studies. They can be cooled with fewer environmental complications than micelle-encapsulated nanotubes, such as the freezing of the suspension. Also of importance to this work, the ambient atmosphere for pillar-suspended nanotubes can be controlled with great flexibility. In this paper, temperature dependent PL measurements on both ensembles and individual nanotubes are reported for seven different assigned SWNT species. In all cases, the PL peak emission energy depends only weakly on temperature between 300 K and 5 K. The integrated intensity and linewidth also change moderately. In addition to PL peaks assigned to particular SWNT species, we report on new spectral features present in intermediate and low temperature spectra.

## II. EXPERIMENTAL DETAILS

The suspended SWNTs in this work were synthesized by chemical vapor deposition, as described in detail previously.[1,9] Measurements were performed on samples grown using methane with either Co or Fe catalysts on silicon substrates. The silicon substrates were patterned with square arrays of silicon or silicon dioxide pillars about 0.5 μm high, with varying spacing between 0.4 and 1 μm. The larger the spacing between pillars, the less likely it is that a nanotube will bridge them.[9] For ensemble studies, samples with narrow pillar spacings and thus a high yield of suspended SWNTs (one or more suspended nanotubes/pillar pair) were used. These are thus relatively small ensembles, with perhaps thousands of suspended nanotubes in an ensemble. Note that in addition to the suspended nanotubes, nanotubes also grew on the flat surface below, but we have only been able to detect PL from suspended nanotubes.[1] For single nanotube studies, samples with larger pillar separations and thus lower yields of suspended SWNTs were used. For both ensemble and single nanotube studies, the excitation spot was 10 μm to 20 μm in diameter. The temperature dependent PL measurements were performed in an open continuous flow cryostat, with SWNTs exposed to helium gas ambient.

The choice of helium ambient proved to be essential. Measurements performed in high vacuum (~$10^{-5}$ torr of air) were similar until ~40K when the luminescence intensity rapidly dropped off to the point of being undetectable. If the temperature was quickly ramped directly to 5K, PL could be obtained at first, but in this case it too gradually decayed away. In comparison, nanotubes cooled in helium ambient remained luminescent at all temperatures and no degradation was observed over many hours of the measurement.

The quenching effect can be attributed to molecular adsorption of atmospheric gases on the nanotube walls. The vapor pressure of $N_2$, $O_2$, and Ar as well as some other trace gases are all of order $10^{-5}$ torr at temperatures around 40 K. The vapor pressure of $H_2O$ is far lower, dropping to $10^{-5}$ torr well above 100 K. Thus it is expected that these atmospheric gases would condense out onto any cold surfaces through cryopumping at these temperatures. It is worth noting that under the assumption of perfect sticking the time to form one complete monolayer on a pristine surface exposed to $10^{-5}$ torr is of order 1 s. In comparison, helium has a vapor pressure of close to an atmosphere at 40K, and the vapor pressure remains above 10 torr even at 5 K.[10] It is also chemically inert. Thus it is an excellent choice of ambient, and is more economical than ultra-high vacuum (UHV) experiments, which could otherwise be used to eliminate any significant adsorption.

Temperature dependent PL and PLE measurements in helium ambient were performed in a continuous flow cryostat. In order to minimize the excitation spot size and maximize collection efficiency, an aspheric lens (4 mm focal length, 0.55 NA) was mounted directly inside the cryostat, with excitation and collection through the one aspheric lens. An external lens was far less efficient owing to the larger working distance and smaller numerical aperture possible. For excitation two different lasers were used in continuous wave mode, a HeNe laser (1.959 eV [632.8 nm]) and a tunable Ti:sapphire laser (1.482 eV [837 nm] to 1.710 eV [725 nm]) with excitation power between 0.3 mW and 0.5 mW. This was focused to a 10 μm to 20 μm diameter spot, which overlapped of order hundreds to thousands of pillars on the substrate. While laser heating may be important at higher power densities, a simple calculation shows that for the current experimental conditions, the effect should be minimal. In the extreme limiting case of a nanotube absorbing all photons within one micron and losing heat only through its ends, the nanotube temperature should increase by no more than a few Kelvin. The luminescence was dispersed by a single grating spectrometer (149 grooves/mm with ~1 meV resolution, or 600 grooves/mm with ~250 µeV resolution) onto a liquid nitrogen cooled InGaAs 512 photodiode array with sensitivity from visible to 1650 nm (0.75 eV). The detector accumulation times were typically between 3 and 30 sec.

## III. RESULTS and DISCUSSION

### 1. Ensemble measurements

Figure 1 shows temperature dependent PL spectra obtained from a nanotube ensemble for excitation at 1.959 eV (632.8 nm) and 1.621 eV (765 nm). For excitation at 1.959 eV (Fig. 1(a)), the spectrum at 260 K is dominated by the (7,5) and (7,6) nanotube species with emission at 1.129 eV and 1.232 eV, respectively. Here, and throughout this work, the (*n,m*) assignment is made by choosing the closest peak in the PLE map to the scheme of Refs. [3] and [11], taking into account the small redshift of those micelle-encapsulated samples compared to our pillar suspended nanotubes.[6] For both the (7,5) and (7,6) species, the optical absorption at $E_{22}$ is peaked around 1.923 eV (645 nm).[3,11] These species seem particularly abundant in one specific sample grown using the cobalt catalyst.

Other nanotube species contribute to the broad emission between 0.8 eV and 1.1 eV. The small features at 0.901, 1.001, 1.028, and 1.076 eV, each labeled with an asterisk only, most likely arise from nanotubes with $E_{22}$ close to resonance with the 1.959 eV excitation. Reasonable assignments for these peaks are (12,2), (11,1), (10,3) and (9,5) respectively. However, the PLE spectroscopy needed to confirm the assignment was not performed, and thus the detailed temperature dependence of these species is not presented here.

Figure 1(b) shows the same kind of spectra with excitation at 1.621 eV (765 nm). In this case, the 280 K spectrum is dominated by three peaks previously assigned to (9,8), (9,7), and (11,3) nanotube species with emission at 0.901, 0.958, and 1.006 eV, respectively.[6] Other peaks labeled "MT", emerging as the temperature is reduced, are previously unreported and will be discussed in detail below.

Between 300 K and 20 K all five assigned nanotube species clearly identified in Fig. 1(a) and (b) show very little or no shift of the emission energy. In Fig. 2, the shift in emission energy is plotted and reveals a monotonic blueshift when the temperature is lowered from 300 K to 20 K. The magnitude of the shift depends strongly on the nanotube species. For the (7,6) nanotube, the total shift in emission energy is 9 meV while for the (11,3) nanotube, it is less than 1 meV. The maximum observed shift was ≈0.9% upon cooling from room temperature to 5K. Defining the PL bandgap temperature coefficient as

$$\kappa_{PL} = \frac{1}{E_{11}} \frac{dE_{11}}{dT}\bigg|_{300K},$$

the largest magnitude coefficient measured is $\kappa_{PL} \approx -3\times10^{-5}$ K$^{-1}$. It is not yet clear how to correlate the change of $E_{11}$ with the particular nanotube species.

In conventional semiconductors, two factors contribute to $\kappa_{PL}$: temperature dependent lattice parameters, and bandgap renormalization due to electron-phonon interactions.[12] For conventional semiconductors at low temperatures, the latter effect usually dominates. Raman experiments suggest that the former may also be important for SWNTs. In Raman, relative shifts of the radial breathing mode energy ($E_{RBM}$) were reported with analogous temperature coefficients in the range $\kappa_{RBM} = -2.5\times10^{-5}$ K$^{-1}$ to $\kappa_{RBM} = -7.5\times10^{-5}$ K$^{-1}$,[13,14] quite consistent with the values measured here for $\kappa_{PL}$. However, the connection between $\kappa_{PL}$ and $\kappa_{RBM}$ is somewhat indirect. In addition, bundling effects complicate those Raman measurements. Intertube interactions and intratube interactions are thought to be of roughly equal importance, so softening due to intertube interactions is expected to increase $\kappa_{RBM}$ by about a factor of two compared to isolated nanotubes.[14]

At first, it might be expected that $\kappa_{PL}$ (and $\kappa_{RBM}$) can be explained by a temperature dependent diameter change. However, molecular dynamics simulations predict almost no change in nanotube diameter with temperature,[14,15] therefore the magnitude of electron (or hole) transverse momenta is expected to remain unchanged. Regardless of whether the nanotube diameter remains unchanged or not, the carbon-carbon bond length increases with increasing temperature. As a result, the nearest neighbor C-C overlap integral ($\gamma_o$) is reduced and in the zone folding approximation[16] this leads directly to a reduction of $E_{11}$. Such a picture explains the sign and the smallness of $\kappa_{PL}$, however it does not readily account for the strong (*n,m*) dependence observed here. Thus it is expected that other contributions to $\kappa_{PL}$, such as electron-phonon interactions, may be just as significant. For the (11,3) and (9,7) nanotubes, various contributions must cancel to give the nearly temperature independent bandgap observed.

Figure 3 shows the PL integrated intensity and the full width at half maximum (FWHM) versus temperature for the five SWNT species assigned in Fig. 1. The integrated intensity increases linearly as the temperature is reduced from 300 to 30 K, but then levels off and decreases at lower temperatures. The twofold increase and the functional dependence appear roughly species independent. For the FWHM (Fig. 3(b)), the opposite trend is seen, with the width reaching a minimum and increasing again. The inverse relationship between FWHM and integrated PL intensity is a common feature in the temperature-dependent PL of conventional semiconductors. The reduction in FWHM for all five nanotube species is 3 meV to 5 meV, except for the (9,8) nanotube where the change is only about 1 meV.

The relatively weak temperature dependence of the PL integrated intensity here suggests that non-radiative decay may be insignificant. In general, when non-radiative decay channels are important, the PL integrated intensity is usually very strongly temperature dependent. When non-radiative channels are present in conventional semiconductors, and the thermal energy approaches their characteristic energy, the integrated PL intensity shows an activated dependence, falling as the temperature is increased. In contrast, the integrated PL intensity in Fig. 3(a) declines only slowly and linearly.

On the other hand, time resolved studies performed on micelle-encapsulated nanotubes suggest that non-radiative decay is important. The measured ultra-short picosecond radiative lifetime[17] together with the relatively low quantum efficiency[2,5] strongly suggest that non-radiative decay channels are important. It is not yet clear if this result can be extended to suspended nanotubes. In the future, temperature dependent time resolved PL studies will be able to resolve this apparent contradiction, and help clarify some of the carrier relaxation mechanisms, be they interband or intraband, radiative or non-radiative.

There are a few additional observations to be made with regards to linewidth in ensemble measurements. Among the assigned species, the smaller diameter nanotubes generally had larger linewidths. In fact, at room temperature, the smaller was the diameter, the larger was the linewidth. Apart from the (9,8) nanotube, which showed very little variation in linewidth, crossing two other species on the graph, this held true at all temperatures. While three of five nanotube species in Fig. 3(b) have sub-$k_B T$ linewidths at room temperature, below 150 K, they all have linewidths greater than the thermal energy. Finally, even with significantly different FWHM at 300 K, again with the exception of the (9,8) nanotube, four of five nanotube species have a similar functional dependence on temperature.

Going back to Fig. 1(b), in addition to the peaks assigned to specific nanotube species, many other peaks barely apparent at room temperature emerge at lower temperatures. Four such peaks are labeled "MT" (for "moderate temperature") in Fig. 1(b), located at 0.87 eV, 0.93 eV, 0.98 eV, and 1.04 eV. In all cases, the peak emission energy depends only weakly on temperature while the intensity increases dramatically. According to the assignment scheme proposed in Refs. [3,11] there are no SWNT species expected at these energies.

The temperature dependent integrated intensity of two MT peaks is plotted in Fig. 4. Within the scatter of data, they both show the same temperature dependence. The functional form is very simple, falling off exponentially with temperature, with a characteristic temperature of 78 K (6.8 meV). It is important to emphasize that this is not an activation energy, but rather a simple decay constant. A good fit could not be obtained using a single activation energy.

A low temperature PLE map was made in part to examine the excitation energy structure of the MT peaks. A PLE map for an ensemble of nanotubes is shown complete with species assignment in Fig. 5. The position of the four MT peaks labeled in Fig 1 (b) is indicated with an arrow in Fig. 5(b). There is no obvious correlation between MT peaks and chirality assigned peaks.

Several mechanisms can be proposed to explain the origin of MT peaks. Some possibilities include localization, dopant/defect related levels, exciton related levels, and phonon replicas. Localization effects in SWNTs are observed in several transport experiments at low temperatures.[18,19] They are caused by the finite length of the nanotube, or by defect induced barriers along its length. As a result, the nanotubes become effectively strings of quantum dots, with atomic-like energy spectra. The typical energy scale is 0.1 meV to 1 meV for 1 μm to 0.1 μm localization lengths. In order to produce a ~25 meV energy splitting as measured here, localization on a very small length scale of order 10 nm is needed.[20,21] This would correspond to a large number of defects (≈1 defect per 1000 carbon atoms). In addition, a well-defined PL peak in ensemble measurements could only appear if most localized nanotube "segments" had almost exactly the same energy spectrum. Since defects are likely to be distributed randomly along the length of the nanotubes, localization is a very unlikely explanation.

Dopant or defect levels often become apparent in conventional semiconductors at low temperatures, and their PL intensity is normally activated with temperature.[22] Similarly, exciton related levels also show an activated dependence.[22] This picture is not consistent with the functional form in Fig. 4. However, other temperature dependent phenomena besides the activation of these levels may also be at play. We cannot rule out dopant or defect levels as an explanation for MT peaks, but it can be concluded that if these are the mechanisms, more than just simple activation is necessary.

Phonon replicas are also a likely possibility, with radial breathing modes energy scales in the range of 20 to 30 meV. However, a phonon replica of the $E_{11}$ peak is expected to have a similar absorption peak at $E_{22}$. This is clearly not the case in Fig. 5. MT peaks have a broader excitation resonance, with no clear correlation with any assigned nanotube peaks. While not strictly related to phonon replicas, the presence of the 2D+2G mode in PLE maps[6] suggests at the very least that phonons should not be ignored in the context of PL processes in nanotubes. At present, both dopant/defect and phonon replica mechanisms seem reasonable, and more work is needed to rigorously determine the true origin of MT peaks.

**2. Individual SWNT measurements**

Previous publications show that PL spectra from individual SWNTs are greatly simplified compared to ensemble measurements.[7,8] Single SWNT PL is also very useful for temperature dependent measurements. Individual nanotubes can be measured between 300 K and 5 K, and both energy shifts and emission linewidths can be readily measured. Figure 6 shows PL spectra obtained from a (12,2) nanotube excited at 1.959 eV (632.8 nm) in close resonance with $E_{22}$. (This assignment was not confirmed by PLE, however other possibilities are too far off resonance to be reasonable.). Figure 6(a) shows the PL spectra at various temperatures. All spectra have been normalized to a maximum of unity since thermal drifts make absolute intensity data partially unreliable. The spectra are very simple with none of the complications of ensemble measurements. The peak is asymmetric at all temperatures. As the temperature is reduced, the peak narrows and blueshifts. It is striking that the single nanotube peak width becomes considerably narrower than ensembles, and at low temperatures (<30K) the peak splits into a series of peaks. Both these points will be discussed below.

Data compiled from Fig. 6(a) are presented in Fig. 6(b) and 6(c). The blueshift of the emission energy is linear from 300 to 50 K, with a bandgap temperature coefficient $\kappa_{PL} \approx -5 \times 10^{-5}$ K$^{-1}$. At lower temperatures, emission energy levels off, with an overall blueshift of $\approx 7$ meV between 300 K and 5 K. The trend in the single nanotube data is sufficiently clear and free of uncertainty that a fit is meaningful. The dotted line is a fit to the empirical Varshni functional form[23]

$$\Delta E_T = \Delta E_O - \frac{\alpha T^2}{\beta + T}.$$

The Varshni functional form is usually a reasonably good fit for conventional semiconductors. In this case the best fit parameters were $\Delta E_o = 7.5$ meV, $\alpha = 75$ µeV/K, $\beta = 600$ K. The Varshni equation does not explain the drop in emission energy observed below 20 K, but this may simply be a consequence of the appearance of the fine structures at these temperatures. Note that in the case of multiple peaks only the brightest peak position is plotted.

Despite the difficulty of measuring a precise absolute intensity caused by thermal drift, it is possible to make some qualitative statements about single nanotube PL intensities as a function of temperature. For all individual nanotubes measured, the increase of the PL intensity with temperature decrease is somewhat stronger for individual nanotubes than for ensembles. Like ensembles, the single nanotube PL intensity levels off around 30 K, and decreases when the temperature is further reduced. This is consistent with the observed correlation between the increase in PL intensity and the decrease in FWHM. In Fig. 6(c), the FWHM decreases linearly from 9 to 2 meV between 300 and 20 K, a nearly fivefold reduction. Below 20 K, a slight leveling off of the FWHM is observed. All single nanotubes measured at low temperature have FWHM of $\approx 1.5$ meV to 2 meV, a width significantly lower than that for ensembles. It is also worth noting that for both ensembles and individual nanotubes, the FWHM scales linearly with temperature between 300 and 25 K.

In our previous work on PL from individual nanotubes,[7] we pointed out that at 300 K, sub-$k_B T$ linewidths were measured with no significant broadening between individual tube and ensemble measurements. The low temperature result is quite different. At 5K, individual SWNTs have significantly narrower FWHM than nanotube ensembles. Even so, at lower temperatures the FWHM for single SWNTs is still larger than the thermal energy. For an individual nanotube at 5 K, even a 1.5 meV linewidths is about four times larger than the thermal energy ($k_B T = 0.43$ eV at 5.0 K).

Figure 7 provides a clue to the origin of this broadening in ensembles. In Fig. 7, PL spectra from three different individual (9,8) SWNTs at 20 K are shown, along with the corresponding peak for an ensemble of many (9,8) nanotubes. The main emission from the $E_{11}$ transition is peaked at 911 meV, with essentially no change in energy from peak to peak. For individual nanotubes, satellite peaks appear in addition to the main PL peak, and have different relative intensities and spacing. The extra peaks are too widely spaced from the principal (9,8) peak to fit into the second LT classification described in detail below. We have not yet characterized these specific peaks in detail, but they may be MT peaks. For ensemble measurements, they become unresolved and produce an apparent broadening of the main peak. Even without an explicit assignment, this does provide some insight into ensemble broadening. Here, for ensemble measurements, the main broadening mechanism observed is not the result of inhomogeneous broadening, but rather the result of additional peaks emerging at low temperatures and associated with the principal ($n,m$) peak.

The PLE spectrum from individual nanotubes can also be measured between 300 K and 5 K. Figure 8 shows the PL intensity for a (12,1) nanotube as a function of excitation energy around the $E_{22}$ resonance. Both high and low temperature spectra are fit to a Lorentzian function. Any blueshift of $E_{22}$ is sufficiently small compared to the peak width to be insignificant to within the accuracy of measurement. The linewidth of the resonance falls from 25 meV at 295 K to 15 meV at 22 K, a 40% reduction. For the (12,1) tube the overall blueshift of $E_{11}$ is about 5 meV between 300 and 20 K ($\sim -1.2 \times 10^{-5}$ K$^{-1}$).

The PL spectra from individual nanotubes often become rich with additional peaks when the temperature is lowered from 20 to 4 K. The peaks that emerge in this temperature range have very specific characteristics and thus we group them together under the designations "LT" for "low temperature" peaks. These LT peaks are already visible at the lowest temperatures shown earlier in Fig. 6(a). Figure 9(a) is a very good example of the same type of emerging peak structure for a (9,8) nanotube. Such peaks are not resolved in ensemble measurements, likely because of the broadening mechanisms outlined above. The rapid evolution of the spectrum over a fairly limited range of temperatures is striking. At 18K, there is only one emission peak. At 7 K, the spectrum is dramatically changed, with a series of satellite LT peaks on the low energy side of the main (9,8) peak and a smaller satellite on the high energy side. The low energy satellite peaks are evenly spaced, with a typical 2 meV to 3 meV spacing. The energy spacing is consistent with the temperature range for which these peaks are observed (e.g. $k_B T$ for 20K is 1.7 meV). An additional peak at 886 meV is not rigorously assigned, but may simply be another of the MT peaks described above.

A PLE map of these new peaks made at 5K is presented in Fig. 9(b). The brightest peak is the assigned (9,8) transition. The two brightest satellite peaks have precisely the same $E_{22}$ peak excitation energy as the chirality assigned peak. While the nearest satellite appears to have a very similar shape to the (9,8) peak, the next neighbor appears somewhat more distorted. The third nearest neighbor peak is also correlated in excitation energy, but not nearly as clearly. There is also a much fainter high-energy satellite with a correlated excitation structure. The peak at 886 meV is much broader in excitation, having an entirely different character to the narrowly spaced satellites. This is consistent with its assignment as a MT peak.

As was the case for the MT peaks, different mechanisms may give rise to LT peaks, including localization, dopant/defect levels, excitonic states and phonon modes. Although the energy scale for LT peak spacing is reasonable for localization effects, it can be excluded since uniform energy spacing would require evenly distributed defects along the length of the nanotubes. Dopant, defect and exciton related levels cannot be ruled out entirely, but again there is no clear reason why such peaks should be so regularly spaced.

On balance phonon modes seem to provide a very likely explanation. First, while they were excitation modes not phonon replica modes, we have already seen that Raman scattering effects (2D+2G modes) influence the PLE map.[6] Second, in terms of phonons, carbon nanotubes are more like molecules than bulk semiconductors, and there are a whole series of different phonon modes available at various energy scales.[16] Third, phonon replicas are seen in many semiconductors at low temperature and they have many of the attributes seen in Fig. 9. Phonon replicas are evenly spaced in energy. The excitation spectrum is the same for the phonon replica as for the principal "zero-phonon" line. Phonon replicas have peak shapes similar to the principal peak. But foremost, the energy separation of LT peaks fits adequately with low energy squashing modes.[16] We should add that recently, low energy phonon modes have been invoked to explain resonances in charge transport through similarly suspended SWNTs.[24] Pending definitive proof, we tentatively assign these satellites to phonon replicas of the principal chirality assigned transition.

Assuming the identification is correct, it is possible to infer more about the physics of the interactions. The relative strength of the satellites with respect to the zero-phonon line suggests a strong electron/hole-phonon coupling, with a Huang-Rhys factor $S \approx 1$.[25] It should be noted that the peak heights are usually not regular and they do not follow the expected functional form for phonon replicas,

$$I_n = \frac{S^n}{n!} I_0,$$

which relates the intensity ($I_n$) of the $n^{th}$ phonon replica with the intensity of the zero-phonon line ($I_0$).[25] However, the peak heights appear to have some order, as they often fall off in intensity the further separated they are from the principal peak.

# IV. SUMMARY

This first temperature dependent PL study reveals considerable new information about the physics of carbon nanotubes. Importantly, the influence of the gas atmosphere on nanotube luminescence is described for the first time. Adsorbed atmospheric gases can actually quench luminescence when they condense out of the vapor. This effect is significant because in SWNTs every constituent atoms is on the surface.

Chirality assigned fundamental PL peaks ($E_{11}$) and second van Hove absorption peaks ($E_{22}$) are remarkably stable with temperature – an order of magnitude more stable in energy than conventional semiconductor materials. This stability can be traced back to the stiffness of the carbon-carbon bond. The blueshift upon cooling is chirality/species dependent. Other PL properties, namely intensity, emission linewidth, and absorption linewidth vary modestly when SWNTs are cooled. Linewidths are considerably narrower for single nanotubes as compared to ensembles, but the mechanism of ensemble broadening is not simple inhomogeneous broadening.

Two new classes of PL peaks never reported before emerge at low temperatures. MT peaks, barely visible at room temperature grow exponentially as the temperature is reduced. They are separated by 20 meV to 30 meV from the fundamental chirality assigned PL peaks. Plausible explanations for these peaks may be phonon replicas or dopant/defect levels. At very low temperatures (<20K), evenly spaced LT satellite peaks emerge. These are tentatively assigned to phonon replicas, and are consistent in energy scale with squashing modes.

Low temperature PL is proving to be an important new window on nanotube physics, and follow-up studies are clearly needed. No less important than the basic physics, the high stability of the optical transitions as a function of temperature is very promising from the point of view of optoelectronic applications such as lasers. Finally, effects of gas atmosphere on nanotube luminescence should be of interest from the point of view of basic chemistry as well as sensor applications.

# ACKNOWLEDGEMENTS

We thank R. L. Williams and J. M. Fraser for numerous discussions and everyday assistance in the laboratory. Partial financial support was provided by the NEDO International Joint Research Grant Program.

**Correspondence** should be addressed to J.L. (email: jacques.lefebvre@nrc.ca)

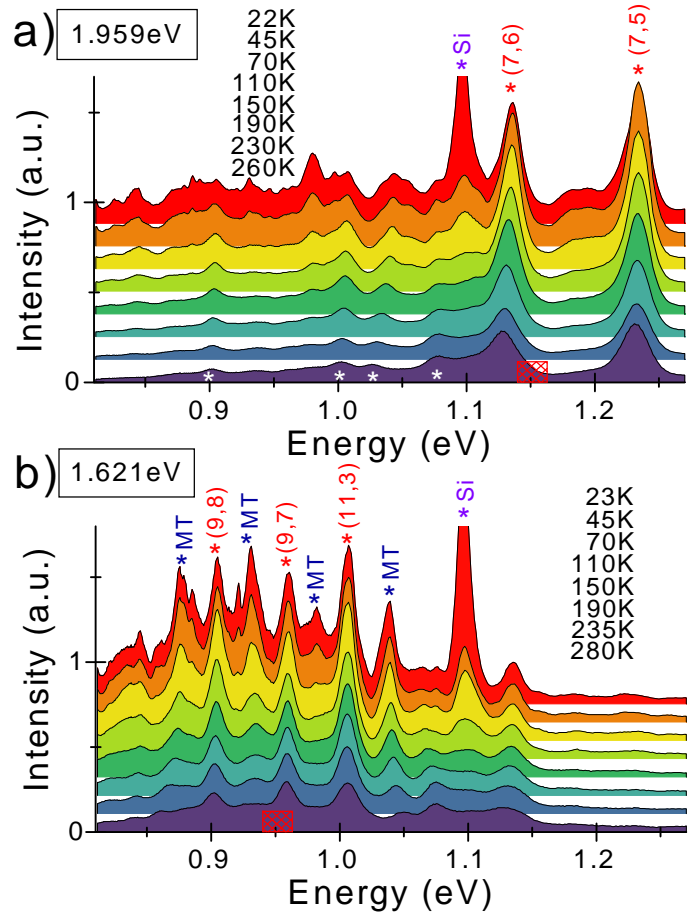

FIG. 1. Temperature dependent photoluminescence spectra from an ensemble of single-walled carbon nanotubes. (a) Excitation at $E_{LASER}$ = 1.959 eV (632.8 nm) (b) Excitation at $E_{LASER}$ = 1.621 eV (765nm). The topmost curve (red) is at the lowest temperature, with temperature increasing as labeled for each successive curve. The peak around 1.1 eV is emission from the Si substrate. Explicitly assigned SWNT peaks are labeled with their (n,m) indices. White stars at the bottom of (a) indicate tentatively assigned (n,m) peaks (see text). Emergent peaks labeled "MT" are discussed in the text. The red hatched squares highlight the average expected position of $E_{11}$ transitions for SWNT species having $E_{22} \approx E_{LASER}$, using an average $E_{22}/E_{11}$ ratio of 1.7.

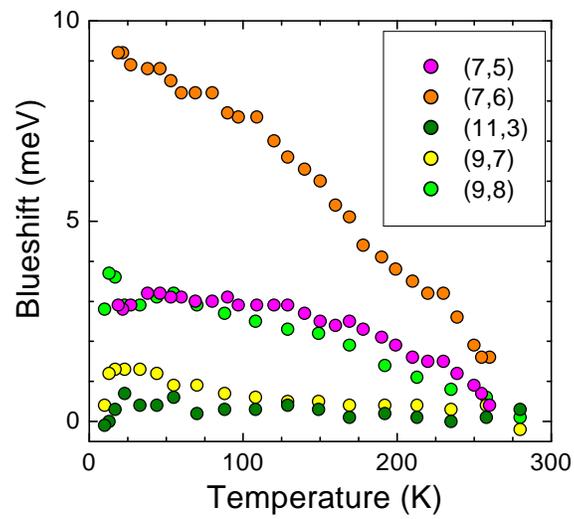

FIG. 2. Blueshift of photoluminescence peak position for an ensemble of single-walled carbon nanotubes. The peak positions were extracted from Fig. 1. The blueshift is defined to be zero at room temperature.

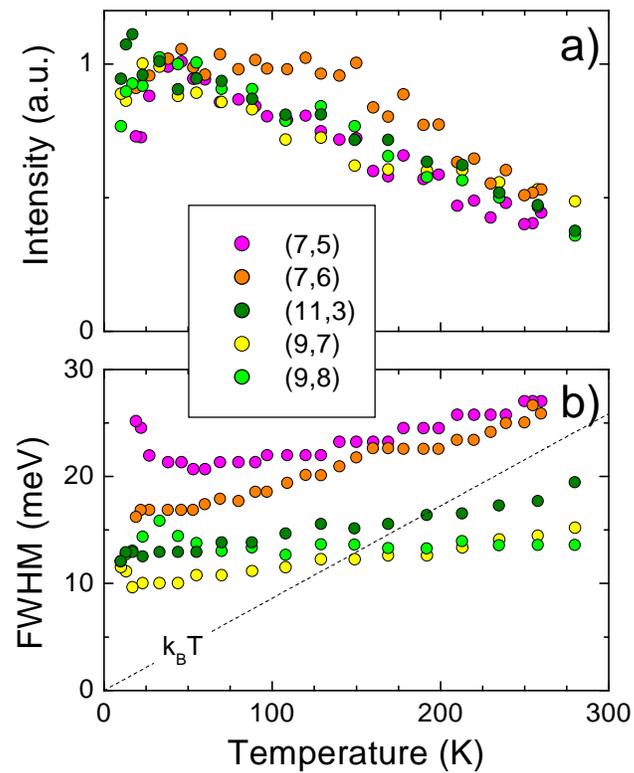

FIG. 3. Temperature dependence of photoluminescence properties for ensembles of five nanotube species. The nanotube species are those assigned in Fig. 1. (a) The integrated PL peak intensity. (b) The PL full width at half maximum (FWHM). The dashed line corresponds to the thermal energy.

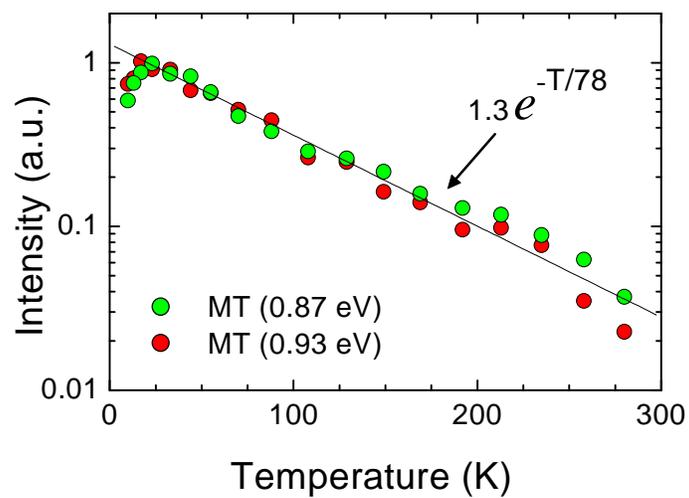

FIG. 4. Temperature dependence of MT peak intensity. The integrated PL peak intensity for two MT peaks is plotted as a function of temperature between 10K to 300 K. The data were extracted from the MT peaks shown in Fig. 1(b).

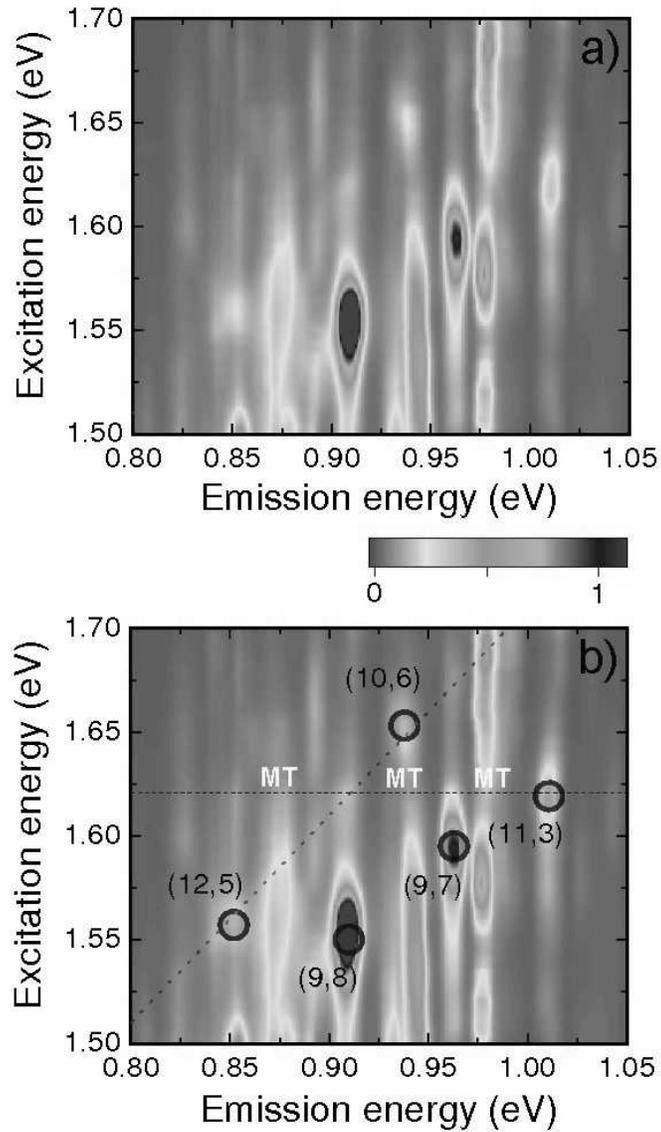

FIG. 5. Low temperature PLE maps. (a) A PLE map for an ensemble of SWNTs at 20K. (b) The same PLE map including peak assignments. Chirality assigned peak positions are labeled with an open circle. The (9,8) and (9,7) species show the brightest PL. The MT peaks are labeled in white. The dotted line overlaps the position of the 2D+2G Raman mode. The horizontal dashed line indicates the excitation energy used for Fig. 1(b).

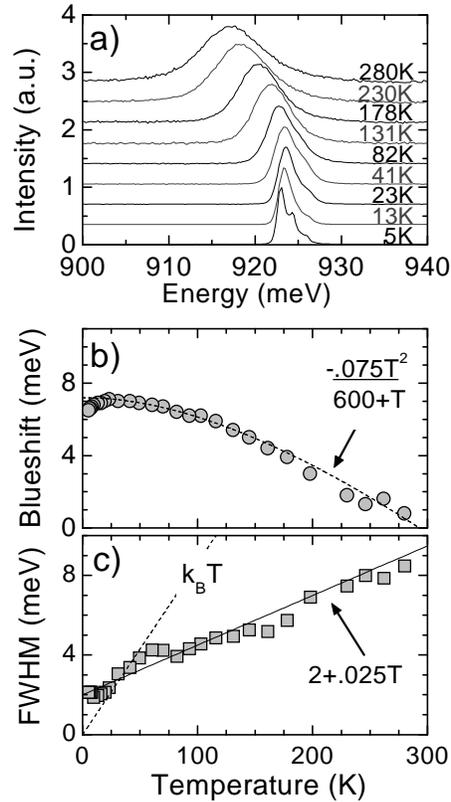

FIG. 6. Temperature-dependent photoluminescence spectra from an individual single-walled carbon nanotube. The laser excitation was at 1.959 eV (632.8 nm). (a) PL spectra at temperatures from 300K to 5K. (b)-(c) Parameters extracted from the spectra displayed in (a). (b) The measured blueshift in emission energy is plotted as a function of temperature, including a fit using the Varshni empirical functional form (dotted line). (c) The measured full width half maximum (FWHM) as a function of temperature with a linear fit (continuous line) and the thermal energy $k_BT$ (dotted line).

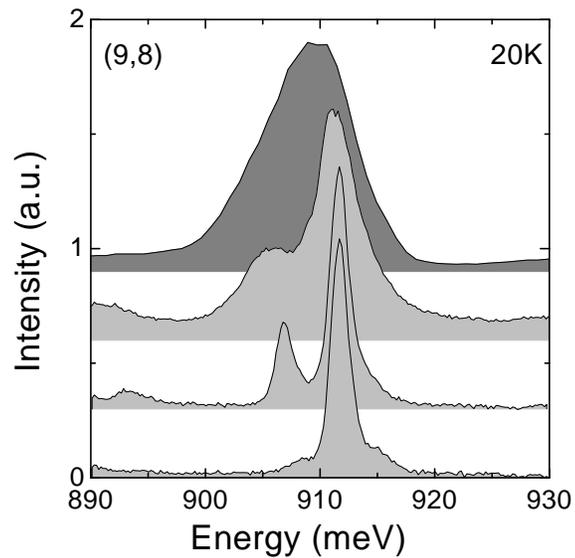

FIG. 7. Comparison of ensemble and individual SWNT photoluminescence spectra. The PL spectra are shown for the (9,8) SWNT species at 20 K, for resonant excitation at 1.55 eV. The dark grey curve is an ensemble of many (9,8) nanotubes. The light gray curves show separate individual (9,8) nanotubes.

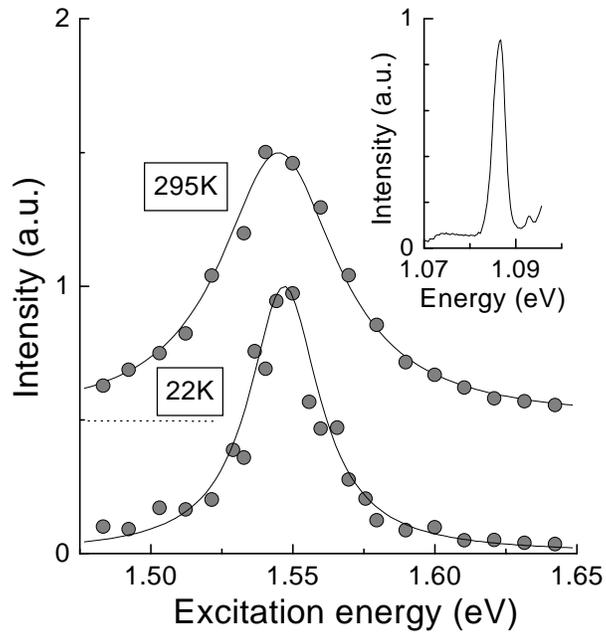

FIG. 8. Photoluminescence intensity as a function of excitation energy for an individual (12,1) nanotube at 295 K and 22 K. The continuous lines are Lorentzian fits to the data, with a 25 meV width at 295 K and a 15 meV width at 22 K. The curve at 295 K is offset and the dotted line indicates the baseline. The inset shows the emission spectrum at 22 K. For reference, the shift in emission energy between 300 and 20 K is about 5 meV.

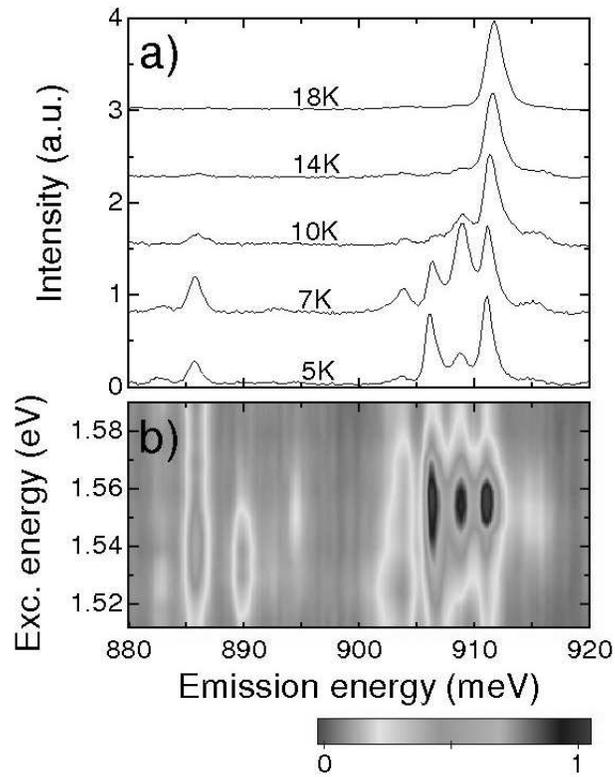

FIG. 9. Low temperature luminescence from an individual (9,8) SWNT. (a) PL spectra are shown at five different temperatures. The laser excitation is at 1.558 eV (796 nm). (b) The corresponding PLE map taken at 5 K.

# REFERENCES

[1] J. Lefebvre, Y. Homma, P. Finnie, Phys. Rev. Lett. **90**, 217401 (2003).

[2] M. J. O'Connell, S. M. Bachilo, C. B. Huffman, V. Moore, M. S. Strano, E. Haroz, K. Rialon, P. J. Boul, W. H. Noon, C. Kittrell, J. Ma, R. H. Hauge, R. E. Smalley, and R. B. Weisman, Science **297,** 593 (2002).

[3] S. M. Bachilo, M. S. Strano, C. Kittrell, R. H. Hauge, R. E. Smalley, and R. B. Weisman, Science **298,** 2361 (2002).

[4] S. Lebedkin, F. Hennrich, T. Skipa, and M.. M. Kappes, J. Phys. Chem. B **107**, 1945 (2003).

[5] S. Lebedkin, K. Arnold, F. Hennrich, R. Krupke, B. Renker, and M. M. Kappes, New J. of Phys. **5**, 140.1 (2003).

[6] J. Lefebvre, J. M. Fraser, Y. Homma, and P. Finnie, cond-mat/0308359; to appear in Appl. Phys. A (2004).

[7] J. Lefebvre, J. M. Fraser, P. Finnie, and Y. Homma, Phys. Rev. B **69**, 075403 (2004).

[8] A. Hartschuh, H. N. Pedrosa, L. Novotny, and T. D. Krauss, Science **301**, 1354 (2003).

[9] Y. Homma, Y. Kobayashi, T. Ogino, T. Yamashita, Appl. Phys. Lett **81,** 2261 (2002).

[10] Vacuum Technology: Its Foundations, Formulae and Tables, Leybold Vacuum Products 1987 p 95.

[11] R. B. Weisman, and S. M. Bachilo, Nanoletters **3**, 1235 (2003).

[12] N. W. Ashcroft and N. D. Mermin, Solid State Physics, Thomson Learning, Inc. (1976) p 566

[13] H. D. Li, K. T. Yue, Z. L. Lian, Y. Zhan, L. X. Zhou, S. L. Zhang, Z. J. Shi, Z. N. Gu, B. B. Liu, R. S. Yang, H. B. Yang, G. T. Zou, Y. Zhang, and S. Iijima, Appl. Phys. Lett. **76**, 2053 (2000).

[14] N. R. Raravikar, P. Keblinski, A. M. Rao, M. S. Dresselhaus, L. S. Schadler, P. M. Ajayan, Phys. Rev. B **66**, 235424 (2002).

[15] Y.-K. Kwon, S. Berber, and D. Tomanek, Phys. Rev. Lett. **92**, 015901 (2004).

[16] R. Saito, G. Dresselhaus, M. S. Dresselhaus, *Physical Properties of Carbon Nanotubes*, World Scientific Publishing Co., Singapore, 1998.

[17] G. N. Ostojic, S. Zaric, J. Kono, M. S. Strano, V. C. Moore, R. H. Hauge, R. E. Smalley, cond-mat/0307154, unpublished.

[18] M. Bockrath, D. H. Cobden, P. L. McEuen, [*] N. G. Chopra, A. Zettl, A. Thess, and R. E. Smalley, Science **275**, 1922 (1997).

[19] A. Bezryadin, A. R. M. Verschueren, S. J. Tans, and C. Dekker, Phys. Rev. Lett. 80, 4036 (1998).

[20] J. Lefebvre, M. Radosavljevic, and A.T. Johnson, Appl. Phys. Lett. **76**, 3828 (2000).

[21] H.W.Ch. Postma, T.F. Teepen, Z. Yao, M. Grifoni, and C. Dekker, Science **293** 76-79 (2001).

[22] J. I. Pankove, Optical Processes in Semiconductors, Dover, New York (1971).

[23] Y. P. Varshni, Physica **34**, 149 (1967).

[24] P. Jarillo-Herrero, S. Sapmaz, L. Gurevich, J. Kong, C. Dekker, L. Kouwenhoven, H. van der Zant, Int. Conf. on the Sci. and Appl. of Nanotubes, Korea (2003).

[25] K. Huang, and A. Rhys, Proc. R. Soc. London, Ser. A **204**, 406 (1950).